\documentclass[useAMS,usenatbib]{mn2e}
\usepackage{aas_macros}
\usepackage{epstopdf}
\usepackage{epsfig}
\usepackage{natbib}
\usepackage{float}
\usepackage{amsmath}
\usepackage{times}
\usepackage{comment}
\usepackage{color}

\title{Examining early-type galaxy scaling relations using simple dynamical models}

\author[Zhang \& Zaritsky]
{Huanian Zhang$^{1}$ and Dennis Zaritsky$^{2}$\\
$^1$Department of Physics, University of Arizona, Tucson, AZ 85721; fantasyzhn@email.arizona.edu\\
$^2$Steward Observatory and Department of Astronomy, University of Arizona, Tucson, AZ, 85721: dennis.zaritsky@gmail.com\\
}

\begin{document}

\date{}

\pagerange{\pageref{firstpage}--\pageref{lastpage}} \pubyear{2015}

\maketitle

\label{firstpage}

\begin{abstract}
We use dynamical models that include bulk rotation, velocity dispersion anisotropy and both stars and dark matter to explore the conditions that give rise to the early-type galaxy scaling relations referred to as the Fundamental Plane (FP) and Manifold (FM). The modelled scaling relations generally match the observed relations and are remarkably robust to all changes allowed within these models. The empirical relationships can fail beyond the parameter ranges where they were calibrated and we discuss the nature of those failures. Because  the location of individual models relative to the FP and FM is sensitive to the adopted physical scaling of the models, unconstrained rescaling produces a much larger scatter about the scaling relations than that observed. We conclude that only certain combinations of scaling values, which define the physical radial and kinematic scale of the model, produce low scatter versions of the FP and FM. These combinations further result in reproducing a condition observed previously for galaxies,  $r_c \rho_0 = $ constant, where $r_c$ is the scaling radius and $\rho_0$ is the central density. As such, we conclude that this empirical finding and global galaxy scaling relations are not independent and that finding the physical cause of one should lead to the solution to the other. Although our models are strictly for  pressure supported galaxies, these results may well hold generally because the central density constraint was first identified in dwarf spheroidals but later extended to rotating giant galaxies and the FM applies to galaxies of any morphological type and luminosity class.
\end{abstract}


\section{Introduction}

We are surprised by the success of galaxy scaling relations, suspecting that Nature deftly executed a subtle swindle. The key is, of course, the nature of our expectations. In truth, those expectations are poorly defined and even less well-justified. We have no {\sl ab initio} analytic physical model of galaxy evolution that works in sufficient detail to provide predictions for scaling relations and numerical simulations of galaxy evolution and formation are only acceptable to us when they reproduce the observed relations. As such, should we be surprised by the scaling relations' success? In this paper, we utilise simple dynamical models of stellar systems, with and without dark matter, to explore the degree to which we should have expected the unexpected and to examine the root cause of the relations.

Among the most stringent of the various galaxy scaling relations
are those that connect galaxies sizes and luminosities to their internal kinematics. Widely used relations include the Tully-Fisher relation \citep[TF;][]{tf} for disk galaxies and the Fundamental Plane \cite[FP;][]{djorgovski, dressler} for spheroidal galaxies. Various attempts to unify and extend these relations exist \citep[for example see][]{burstein}, but we focus here on a straightforward extension of the FP that both unifies these into a single relation and extends the reach of the scaling relation to low mass galaxies and globular clusters \citep[referred to as the Fundamental Manifold, hereafter FM;][]{z06,z07}.  Both the FP and FM relate the size, using the half-light radius ($r_h$), the internal kinematics ($V$), using either or both of the rotational velocity ($v_h$) and line-of-sight velocity dispersion ($\sigma_h$) at the half-light radius, and the stellar concentration, using the surface brightness within the half-light radius ($I_h$).

Because it is possible, with assumptions, to trace the origin of these relations to the Virial Theorem \citep[for one such discussion see][]{z12}, it is natural to suspect that the scaling relations are merely a restatement of the Virial equilibrium of the central portion of galaxies. However,  there are various reasons why the Virial theorem, which is valid for average measurements over the entire system, could grossly fail to predict
the tight empirical sequence of galaxies within the $V-r_h-I_h$ space. Furthermore, galaxies are not structurally homologous \citep[for example, see][]{bertin}, and therefore the scaling relations, and related mass estimators require that any differences among the galaxies, in their structure, dark matter content, or stellar mass-to-light ratio, be either minor or captured by coordinated changes in $V$, $r_h$, or $I_h$. Variations among galaxies in their structure that cannot be captured by a function of $V$, $r_h$, and $I_h$ will result in scatter in the TF, FP, and FM.

 Significant effort has preceded this work in establishing and validating mass estimators, which can be related, but are not identical, to scaling relations \citep{cappellari06,walker,wolf}. To be more specific about the distinction between the two, mass estimators relate a velocity measurement, typically the dispersion, and radius, typically the half-light radius, $r_h$, to the mass. The final estimators are independent of the luminosity or surface brightness of the stellar system, although arriving at a decision on the appropriate radius or aperture for the velocity dispersion measurement is often informed by the
 surface brightness profile \cite[see, for example, ][]{agnello}.  On the other hand, scaling relations such as the FP, TF, and FM may in principle incorporate luminosities or surface brightnesses and  by doing so depend on the efficiency of star formation, the star formation history, mass loss from either stellar or AGN feedback, metallicity, and so forth. It is entirely possible for a stellar system to lie far from standard scaling relations, if, for example, it lost 99\% of its primordial baryons, and for the mass estimator to produce an accurate mass estimate. Scaling relations reflect patterns of evolution beyond simple dynamics.

However, well-known scaling relations fail in certain regimes. For example, the FP cannot be directly extended to low mass systems such as globular clusters and some extreme dwarf galaxies \citep{burstein,forbes,z11}. Numerical simulations appeal to feedback processes to produce dwarf galaxies with realistic properties \citep[see][for some examples]{governato10,sawala10,simpson} and to match the numerical abundance of such systems \citep{sawala}. Even empirical studies, which are based on abundance matching, suggest that the baryonic content of low mass systems is low \citep{papastergis}.  However, the existence of smoothly varying, low scatter, scaling relations argues against strongly stochastic, highly non-linear physics.

Simple dynamical models of stellar clusters and galaxies provide one framework within which to quantitatively investigate how difficult it is to produce tight scaling relations without appealing to baryonic physics. In particular, we desire to develop intuition for how much tuning is necessary to produce a scaling relation of the observed character  within these simple models. Although analytic models are limited in the degree to which they match reality, lacking a connection to cosmologically-linked galaxy evolution, they provide great flexibility and enable us to efficiently probe a wide range of parameter space, including even ``galaxies" with characteristics beyond those found in reality.  Can simple dynamical models reproduce the range of stellar systems we observe, or do scaling relations also point to additional baryonic physics? Is the result of such physics easily described as a simple constraint?

Identifying an appropriate analytic model to use requires
defining a distribution function (DF) that provides a reasonably realistic description of stellar systems. With the DF in hand, we can calculate any observational parameter related to the galaxy's structure and kinematics. However, self consistent models, where the DF defines a mass distribution that gives rise to the gravitational potential used in the DF, are  uncommon and have yet to reproduce the full broad range of features observed in galaxies. Of course, we observe much more detail in galaxies than is considered within scaling relations (for example, the existing scaling relations ignore deviations from elliptical isophotes for ellipticals \citep{bender} and $m \ge 1$ Fourier components in disks \citep{franx,rix}). Perhaps our models do not need to reproduce such features if we use them only to test scaling relations. Instead of demanding that we identify the correct, or even optimal, DF before proceeding, we choose to use the simplest plausible one with sufficient freedom to allow varying the parameters of interest ($v_h$, $\sigma_h$, $r_h$, and $I_h$).

Work along these lines, in particular to produce self-consistent dynamical models that match the observed properties of actual, individual galaxies, exists in the literature. A directly relevant example is the work of \cite{bertin92} and \cite{saglia}. Those authors chose a particular DF that includes rotation, explored the characteristics of those models, and determined whether such models reproduced real galaxies. They find good correspondence between their models, with different dark matter fractions within the luminous regions, and giant galaxies. As such these models must also satisfy the scaling relations. Separately, \cite{mclaughlin} followed an analogous exploration of globular clusters and spherical, isotropic models. However, for our purpose of exploring the full range of galaxy masses from giants to dwarf  spheroidal galaxies and globular clusters, a large range of rotational to pressure support, and the flattening due to velocity dispersion anisotropy, these models do not go far enough. Furthermore, for our goal of exploring the scaling relations we want to also test models that do not match known galaxies. Are the scaling relations something that falls out naturally from the dynamics, whether or not analogous galaxies exist, or must whatever other processes  are responsible for creating galaxies with a limited set of characteristics, such as feedback and self-regulating star formation, also play a key role?

 A critical component of our work here are the models of \cite{wilson}, which are built on a distribution function that includes the component of the angular momentum per unit mass about the symmetry axis, $J_z$, because these models produce systems with  differential rotation and isophote flattening, both of which are present among the full set of early-type galaxies.
By allowing for the possibility of bulk rotation that is radially differential, they provide a distinct advantage over precursor non-rotating King-Michie models \citep{king66} in their approximation of realistic galaxies, but are otherwise simple and have no additional free parameters beyond the two introduced to generate these characteristics\footnote{Wilson introduced two parameters $\beta$ and $\zeta$. Because the current generally accepted description of anisotropy uses the notation $\beta$, we have replaced the letter $\beta$ in the Wilson model with $\chi$. As Wilson explained, $\zeta$ is included to introduce differential rotation in a way that makes it independent of the energy cutoff.}. Wilson noted that his model could easily be extended to include multiple stellar populations. He envisioned that such models would be useful in exploring age or metallicity gradients in early-type galaxies. However, rather than modelling multiple {\sl stellar} populations, we use this idea to include a dark matter component. Because DFs depend on the mass density rather than the individual particle masses, our ignorance regarding the nature of the dark matter particle is not a weakness here. We appreciate that the Wilson DF does not reproduce many key features of early type galaxies, such as triaxiality and figure rotation, and that more sophisticated attempts at deriving a more accurate DF exist \citep{deZ}; however, any increase in sophistication beyond the Wilson DF necessarily involves additional parameters and complexity for which the scaling relations provide insufficient constraint. We opt to explore the properties of this simplest of models to develop a rudimentary intuition regarding the scaling relations and identify which aspects of the model are critical to the existence of scaling relations and which are not. In \S \ref{sec:wilson}, we summarise the principal features of the model and fill in where necessary what Wilson did not present regarding the extension to multiple components. In \S \ref{sec:results} we present results from models that only have a stellar component (\S \ref{sec:single}) and then from models with both stars and dark matter (\S \ref{sec:double}). We conclude by summarising our findings in \S \ref{sec:summary}.

\section{Development and application of Wilson's model}
\label{sec:wilson}

The application of the Wilson models requires
that the gravitational potential be time independent and azimuthally symmetric. Therefore, two of the integrals of motion are $E$, the energy per unit mass, and $J_z$, the angular  momentum about the rotation axis per unit mass. A valid distribution function is then constructed as a function of the kinetic energy, the potential energy, and the projection of the angular momentum onto the rotation axis.
In spherical coordinates, (r, $\theta$, $\phi$), with the corresponding velocities $v_r$, $v_\theta$, $v_\phi$, then,

\begin{equation}
\begin{split}
E &= \frac{1}{2}(v_r^2 + v_\theta^2 + v_\phi^2) + U(r,\theta),\\
J_z& = r (\sin\theta)v_\phi
\end{split}
\end{equation}

\noindent
where $U$ is the gravitational potential. The DF adopted here  is the extension Wilson suggested of his single component DF to $i$ components,
\begin{equation}
\begin{split}
{\rm DF}(E,J_z) = \sum\limits_{i}n_i[{\rm exp}(-\alpha_iE)-1+\alpha_iE] \times \\
 {\rm exp} (\chi_iJ_z-\frac{1}{2}\zeta_i^2J_z^2)
 \end{split}
\label{eq:DF}
\end{equation}
\noindent
where $\chi_i$ and $\zeta_i$ set the angular velocity profile for the $i$th component, $\alpha_i$ sets the relative energy of the components, and $n_i$ sets the relative densities.
For a given choice of parameter values,
the model is solved using the implicit iterative approach described by Wilson, although the numerics we use are in detail different. In principle, we solve Poisson's equation and self consistently define the density field that results in the potential required to define the DF.
 This is a well established approach that predates Wilson's study \citep[]{1970AJ.....75..674P}.

For clarity, we reprise some of the material from \cite{wilson}. The scaling of physical quantities into dimensionless ones as used in the calculations is achieved by defining

\begin{equation}
r_* = r_c\times r, \qquad U_* = v_*^2 \times U, \qquad \rho_* = \rho_{0*}/\rho_0 \times \rho
\end{equation}

\noindent
where the physical variables are distinguished from the dimensionless quantities by asterisks,
$\rho_0$ is the central density, the relation between $\rho_{0*}$, the scaling radius, $r_c$, and the scaling velocity, $v_*$, is given as

\begin{equation}
\rho_{0*} = \frac{9}{4\pi G}\frac{v_*^2}{r_c^2}.
\end{equation}

Likewise, the DF parameters, $\alpha_i$, $\chi_i$, and $\zeta_i$, in principle have dimensioned (starred) counterparts to match their dimensioned associated quantities, $E$, $J_z$, and $J_z^2$ respectively,  but they are dimensionless in Eq. 5 and beyond where we have transformed all physical quantities to dimensionless versions. Because we never compare the values of 
$\alpha_i$, $\chi_i$, and $\zeta_i$ to observables, the existence of dimensioned versions is of academic interest and we will simply quote the numerical values of the dimensionless versions. On the other hand, to obtain estimates of densities, masses, energies, and so forth we need to transform those quantities back to their dimensioned counterparts.

As always, moments of the DF result in different measurable descriptions of the system, such as the stellar density or moments of the velocity field. Wilson presented the relevant equations for a single component model, but left the calculation of the corresponding multi-component equations as an exercise for the reader. Here we provide the missing equations.
The density of the $i$th component, calculated by integrating the distribution function over the velocities, is
\begin{equation}
\begin{split}
\rho_i(r\sin\theta,U) = 2\pi n_i  \int_{-v_e}^{v_e}  \exp(\chi_i r\sin\theta v_\phi-\frac{1}{2}\zeta_i^2r^2\sin^2\theta
       v_\phi^2) \times \\
(\frac{1}{\alpha_i} \exp({-\alpha_i\varepsilon})-\frac{1}{\alpha_i} + \varepsilon -\frac
 {\alpha_i \varepsilon^2}{2})\ dv_\phi,
 \end{split}
\label{eq:density}
\end{equation}
\noindent
where  $\varepsilon = U + \frac{1}{2}v_\phi^2$, and $v_e = \sqrt{(-2U)}$, the latter being the escape velocity. We obtain the moments of the velocity distribution of the $i$th component by integrating the DF over the corresponding velocities,
\begin{equation}
\begin{split}
\langle v_{\phi} \rangle_i =& \ \  \rho_i^{-1}\iiint\limits_{v_r^2+v_\theta^2+v_\phi^2 \le v_e^2}n_i[{\rm exp}(-\alpha_iE)-1+\alpha_iE] \times \\
 & {\rm exp}(\chi_iJ_z-\frac{1}{2}\zeta_i^2J_z^2) v_{\phi} \ dv_r dv_\theta dv_\phi \\
\\
=&\ \ 2\pi\rho_i^{-1}n_i\int_{-v_e}^{v_e} \exp(\chi_i r\sin\theta v_\phi-\frac{1}{2}\zeta_i^2r^2\sin^2\theta v_\phi^2) \times \\
& (\frac{1}{\alpha_i} \exp({-\alpha_i\varepsilon})-\frac{1}{\alpha_i} + \varepsilon -\frac
 {\alpha_i \varepsilon^2}{2}) v_{\phi}\ dv_\phi\\
\end{split}
\end{equation}

\begin{equation}
\begin{split}
\langle v_{\phi}^2 \rangle_i =& \ \  \rho_i^{-1}\iiint\limits_{v_r^2+v_\theta^2+v_\phi^2 \le v_e^2}n_i[{\rm exp}(-\alpha_iE)-1+\alpha_iE] \times \\
& {\rm exp}(\chi_iJ_z-\frac{1}{2}\zeta_i^2J_z^2) v_{\phi}^2 \ dv_r dv_\theta dv_\phi \\
\\
=&\ \ 2\pi\rho_i^{-1}n_i\int_{-v_e}^{v_e} \exp(\chi_i r\sin\theta v_\phi-\frac{1}{2}\zeta_i^2r^2\sin^2\theta v_\phi^2) \times \\
& (\frac{1}{\alpha_i} \exp({-\alpha_i\varepsilon})-\frac{1}{\alpha_i} + \varepsilon -\frac
 {\alpha_i \varepsilon^2}{2})v_{\phi}^2\ dv_\phi\\
\end{split}
\end{equation}

\begin{equation}
\begin{split}
\langle v_r^2 \rangle_i =& \ \  \rho_i^{-1}\iiint\limits_{v_r^2+v_\theta^2+v_\phi^2 \le v_e^2}n_i[{\rm exp}(-\alpha_iE)-1+\alpha_iE] \times \\
& {\rm exp}(\chi_iJ_z-\frac{1}{2}\zeta_i^2J_z^2) v_r^2 \ dv_r dv_\theta dv_\phi \\
\\
=&\ \ 2\pi\rho_i^{-1}n_i\int_{-v_e}^{v_e} \exp(\chi_i r\sin\theta v_\phi-\frac{1}{2}\zeta_i^2r^2\sin^2\theta v_\phi^2) \times \\
& (\frac{1}{\alpha_i^2} \exp({-\alpha_i\varepsilon}) - \frac{1}{\alpha_i^2} + \frac{\varepsilon}{\alpha_i} -\frac{\varepsilon^2}{2} + \frac{\alpha_i\varepsilon^3}{6})\ dv_\phi\\
\end{split}
\end{equation}
\noindent
Then the projected mean rotational velocity $v_r$ and velocity dispersion $\sigma$ can be calculated using the velocity moments  (from \citep{wilson}),

\begin{equation}
v_r = \Sigma^{-1} \int_{-\infty}^{+\infty} \nu \langle v_{\phi} \rangle \rho \ dx
\end{equation}

\begin{equation}
\sigma^2 = \Sigma^{-1} \int_{-\infty}^{+\infty} [(1 - \nu^2) \langle v_r^2 \rangle + \nu^2 \langle v_{\phi}^2 \rangle] \rho \ dx  - v_r^2.
\end{equation}

\noindent
 Where $\nu$ is the geometric factor by which the velocities are projected onto the line of sight, $\Sigma$ is the projected surface density, defined as $\Sigma = \int_{-\infty}^{+\infty}\rho \ dx$, where $x$ represents distance along the line of sight.
To evaluate quantities in physical units, again following \cite{wilson}, we express the central surface density, total mass, total potential energy, and total kinetic energy as

\begin{equation}
\Sigma_{0*} = r_c \rho_{0*} / \rho_{0} \int_{-\infty}^{+\infty}\rho \ dx
\end{equation}
\begin{equation}
m_* = \frac{r_c v^2_*}{G} m, \qquad W_* = \frac{r_c v^4_*}{G} W, \qquad K_* = \frac{r_c v^4_*}{G} K.
\end{equation}

To explore how our models reproduce the FP and FM we must calculate the following quantities from the models: the projected half  stellar mass radius\footnote{We calculate the projected half light or stellar mass radius using circular apertures. This choice is often adopted in empirical studies, as done as a baseline approach by \citet{2013MNRAS.432.1709C}. Those authors, however, did a careful examination of how various possible definitions of the scaling radius relate to the more physical, gravitational radius. They conclude that resulting differences are relatively minor and explain why the classical empirical choice of the effective radius yields robust scaling relations.}, which corresponds to the half light radius because of our adopted constant value of the stellar mass-to-light ratio, $\Upsilon_*$, the projected line-of-sight stellar velocity dispersion,  we choose\footnote{ Observational studies of the FP typically rescale their observed integrated velocity dispersions to a fixed radius of $r_e/8$ using a correction as advocated by \cite{jorgensen}. For our adopted definition this correction would simply rescale all $\sigma$ by a constant, thereby affecting the normalization of the relationships. Because we already have freedom to renormalize the relationships due to the unknown value of $\Upsilon$ we neglect this rescaling. For observational studies this scaling is more critical because the effective aperture of each observation is different in units of $r_e$ and so the scaling is needed to place the measurements on an equal footing.} to evaluate the average value within $r_h$, the stellar rotation velocity, $v_r$, again averaged within $r_h$  which we refer to as $v_h$, and the mass of the stellar component projected interior to $r_h$, from which we can calculate $I_h$ using $\Upsilon_*$. For models that include dark matter, we will also calculate the dark matter mass projected interior to $r_h$, which we need in combination with the stellar mass to calculate the total mass-to-light ratio within $r_h$, $\Upsilon_h$.

The form of the well-known FP \citep{djorgovski,dressler} is
\begin{equation}
\log{r_h} = A\log{\sigma_h} - B\log{I_h} + C_{FP},
\end{equation}
\noindent
where the coefficients, $A$, $B$, and $C_{FP}$, are empirically fit. We adopt the values of those coefficients presented by a recent determination \citep[$A = 1.49$, $B=0.75$;][]{bernardi}.  The constant $C_{FP}$ is not of particular relevance here because we are free to choose the mass-to-light ratio for our models, which corresponds to freedom in setting $C_{FP}$. This freedom exists throughout this study and so we are free to shift any of our models vertically in our plots. Purely translational discrepancies between the models and the data, to the degree that they can be accounted for by a plausible change in the mass-to-light ratio, are meaningless.  For models including only stars, this constraint is fairly stringent and set by stellar population models, but for models that include dark matter a wide range of values,  corresponding to variations in total M/L from values of a few to $\sim$ 100 at least \citep{mateo}, is plausible.
In comparison, the FM is given by
\begin{equation}
\log{r_h} = 2\log{V} - \log{I_h} - \log{\Upsilon_h} + C_{FM}
\label{equation:FM}
\end{equation}
\noindent
where $V \equiv \sqrt{(\sigma_h^2 + v_h^2/2)}$,  the dependence on the mass-to-light ratio is made explicit,  and $C_{FM}$ is a constant \citep{z07}. Again, translations of the models can be affected through the choice of $\Upsilon$.

Although we can calculate $\Upsilon_h$ in our models, in practice $\Upsilon_h$ is generally not known or easily measurable. Variations in $\Upsilon_h$ among galaxies have either been implicitly fitted out, as done when deriving the coefficients of the FP, or are accounted for using an empirical fitting formula for $\Upsilon_h$, $\Upsilon_{fit}$, as done for the FM \citep{z12}. The principal assumption in either case is that variations in $\Upsilon_h$ are sufficiently well described using a function that depends only on $V$, $r_h$, and $I_h$. When placing the models calculated here on the FM, we will take two approaches. First, we will use the $\Upsilon_h$ evaluated directly from our models (for an adopted $\Upsilon_*$) and we refer to this version of the FM as the theoretical FM (TFM). Second, we will use the $\Upsilon_h$ estimated by calculating $\Upsilon_{fit}$ and we refer to this version of the FM as the empirical FM (EFM). We adopt the empirical fitting function for $\Upsilon_{fit}$ provided by \cite{z12} and do not recalibrate it from our models.
The use of the two different approaches helps us examine the degree to which the empirical fitting function happened to remove not only variations in $\Upsilon_h$ but also to inadvertently remove variations in structure that can be tracked using $V$, $r_h$ or $I_h$.

\section{Results}
\label{sec:results}

We structure our discussion of the results by stepping through them, beginning with the single component (stellar) models. We discuss the role of various parameters with respect to the placement of stellar systems on the FP and FM. Then we progress to explore the two-component (stellar + dark matter) models, again focusing on the role of various parameters on the placement of systems on the FP and FM. Particularly for the star + dark matter models, we compare the properties of our model systems to the broad characteristics of galaxies to ensure that we are examining analogous systems, but we do not attempt to match specific galaxies. For the FM, we present results using both the model calculated value of $\Upsilon_h$, the TFM, and the empirical value obtained using the fitting function, the EFM. To the degree that the results are the same from these two approaches, that agreement confirms the interpretation of the empirical fitting function as descriptive of $\Upsilon_h$. To the degree that the results differ, that disagreement highlights other systematic behaviour, for example in the spatial structure, that can be described as a function of $V$, $r_h$, and $I_h$ and that has been unwittingly captured within the empirical fitting function.

 It is important to appreciate that we are dealing with two kinds of adjustable parameters in our models. The first are those directly related to physical quantities, such as the scaling radius, $r_c$. The scaling between the dimensionless versions of these parameters used in the calculations and the dimensional versions of the real world have been presented in the previous section. However, changes in these quantities mean that the systems we are modeling can span a range of different real-world systems, from dwarf spheroidal galaxies to giant ellipticals. Such changes, in a cosmological simulation of galaxy formation, will also correspond to different evolutionary histories, environments, and could span a range of masses where different baryonic processes, such as stellar feedback vs. AGN feedback, are key. The second are those related to the internal structure of the model ($\alpha$, $\chi$, $\zeta$). Altering these affects the kinematic and spatial structure of the stellar system, but to first order such models represent the same type of galaxy. Even though choice of $\chi$ and $\zeta$ affect the rotational properties and flattening of the model, these models are not able to represent disk galaxies and so we are limited to early-type galaxies. 

\subsection{Models with stars only}
\label{sec:single}
\subsubsection{Spherical and isotropic velocity dispersion}

\begin{figure}
\begin{center}
\includegraphics[width = 0.48 \textwidth]{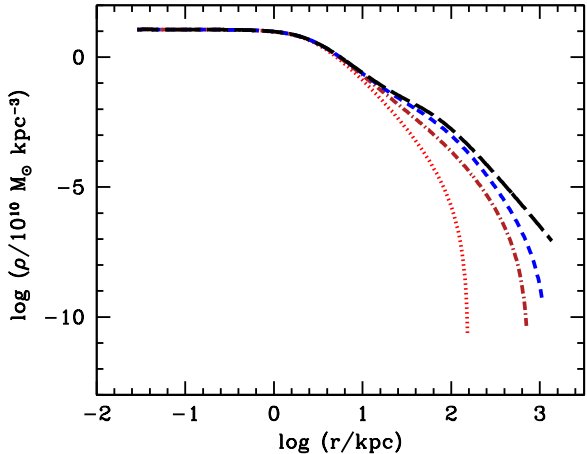}
\end{center}
\caption{Relative density profiles for models with the rotational parameters of Wilson's (1975) Model A ($\chi=0.133$ and $\zeta=0.067$), $r_c = $ 2 kpc and varying central potential. Lines represent models with $U_0 = -6$ (dots), $-8$ (dot-dash),  $-10$ (short-dash), and $-12$ (long-dash) and illustrate the change in behaviour with changing $U_0$. }
\label{Figure:density_non}
\end{figure}

We begin by discussing  models that are spherical and have isotropic velocity dispersions obtained by setting $\chi$ and $\zeta$ to be 0.
These models are spherical, tidally truncated models of the King-Michie type.  For completeness, we note that we set $\alpha_1$ and $n_1$ to 1 in these single component models.
As we make the central potential, $U_0$, deeper (more negative), the profile progresses from that representative of a globular cluster, with a relatively lower concentration and sharper tidal boundary, to that more representative of elliptical galaxies, with something reminiscent of a de Vaucouleur's r$^{1/4}$ profile.
The density profiles for different central potentials are shown in Figure~\ref{Figure:density_non}. Independent of $U_0$, the density remains roughly constant for the smallest radii. 
This plot reproduces results known since the work of \cite{king62} and is only provided for reference.

\begin{figure}
\begin{center}
\includegraphics[width = 0.7 \textwidth]{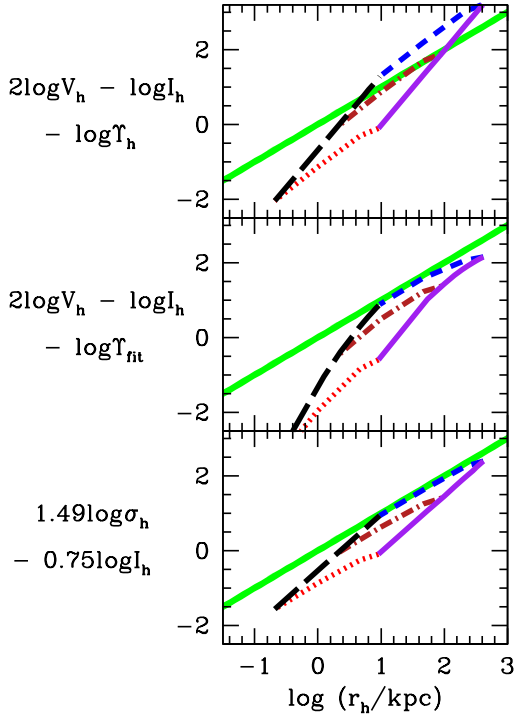}
\end{center}
\caption{Isotropic, non-rotating stellar models relative to the FM and FP. We plot their location relative to the TFM (using the known model value of $\Upsilon_h$) in the top panel,  the EFM (using the empirical fitting function for $\Upsilon_{h}$, $\Upsilon_{fit}$) in the middle panel, and the FP in the bottom panel. Model tracks represent results for $U_0 = -6$ (black long dash),  $U_0 = -12$ (purple solid) as we vary $r_c$ from 0.1 to 4.6 kpc and for $r_c$ = 0.1 kpc (red dots), $r_c$ = 1 kpc (dark red dot-dash) and $r_c =$ 4.6 kpc (blue short dash) as we vary $U_0$ from $-6$ to $-12$.
The 1:1 line represents the observed scaling relation in these axes and the width of the line matches the typically observed 0.1 dex scatter.   Because the adopted $\Upsilon_*$ is arbitrary, we set $\Upsilon_*$ to a plausible value of 1. The models may be shifted vertically, by changing the adopted $\Upsilon_*$, to produce an optimal match.}
\label{Figure:rc_vary}
\end{figure}

We now calculate the properties of these models varying both $U_0$ (from $-12$ to $-6$) and $r_c$ (from 0.1 to 4.6 kpc, broadly representing all early-type galaxies from dwarf spheroidal, to compact elliptical, to giant elliptical) and place them on the FP and FM  in Figure~\ref{Figure:rc_vary}.
The results illustrate several key points. First, an observed range in $r_h$ among a set of stellar systems can reflect either a change in the scaling radius, $r_c$, for a fixed potential or  a change in the central potential for a fixed $r_c$ (combinations of the two are also allowed). Galaxies of different sizes along a scaling relation can therefore, in principle, be either larger (or smaller) versions of the same class of object, that is with the same $U_0$ and hence similar density profile, but with different $r_c$, or galaxies with different concentrations, as reflected by varying values of $U_0$, but with similar $r_c$.  We step through our results for completeness, although many other studies have reflected on many of the same points in other contexts \cite[see for example,][]{saglia}.

Second, these simple models easily fill portions of parameter space that are beyond those populated by real galaxies. Therefore, arbitrary combinations of $r_c$ and $U_0$ are not created in nature.
No single set of models with a fixed central potential depth can explain the range of systems along the FP or FM because changes in $r_c$ eventually move galaxies too far from the scaling relation before being able to replicate the range of $r_h$ (recall that vertical shifts are allowed by our freedom in setting $\Upsilon_*$, so we cannot conclude which choice of $r_c$ is ``correct" for a particular choice of $U_0$). To populate the FP or FM with models such as these requires some coordination between $r_c$ and $U_0$. 


Third, the observed scaling relations are well matched by models selected to have different values of $U_0$ and 
 a fixed $r_c$. 
This result will also be confirmed as we progress through the models and suggests that the key to reproducing the galaxy scaling relations lies in constraining the range of the scaling parameter $r_c$.

The models have two scaling parameters, $r_c$ and $v_*$. We just described how varying $r_c$ can help the models cover the parameter space spanned by real galaxies, so one might wonder about the effect of varying $v_*$. Varying $v_*$ enables one to 
reproduce galaxies with a range of velocity dispersions, but it does not result in moving models within the parameter space of Figure \ref{Figure:rc_vary} because resulting changes in the $2\log (V_h)$ term are balanced by changes in the $\log \Upsilon_h$ term. As such, for our
particular purpose of comparing models with the scaling relations, we can fix or vary $v_*$ to any value and not affect our conclusions.

\subsubsection{Rotating models with varying $\chi$ and $\zeta$}
\label{sec:beta_case}

Next we consider the effect of adding rotation, as parametrised by $\chi$ and $\zeta$, on the location of our single population (stellar) models on the FP and FM.  The introduction of nonzero values for these parameters also results in departures from spherical symmetry so viewing angle becomes an additional parameter if one intends to match individual systems. We vary $\chi$ from 0.01 to 0.26, and fix $\zeta = 0.067$ and $r_c= 2$ kpc. We adopt the value of $\zeta$, and inclination (65$^\circ$), to match Wilson's Model A, as plausible values for a realistic galaxy, and arbitrarily set the scaling radius to our selected value. Because $\chi$ is not an easily interpretable parameter, we note that this range of $\chi$ results in a corresponding range of $v_h/\sigma_h$ of 0.135 $-$ 1.20 for $U_0 = -12$, $0.041 - 0.926$ for $U_0 = -10$,  $0.03 - 0.628$ for $U_0 = -8$, and $0.016 - 0.378$ for $U_0 = -6$. As such, these models span the range of $v_h/\sigma_h$ measured for elliptical and S0 galaxies \citep{vdm07,bedregal}, but, importantly, they do not span the same range of $v_h/\sigma_h$ for each choice of $U_0$.

\begin{figure}
\begin{center}
\includegraphics[width = 0.7 \textwidth]{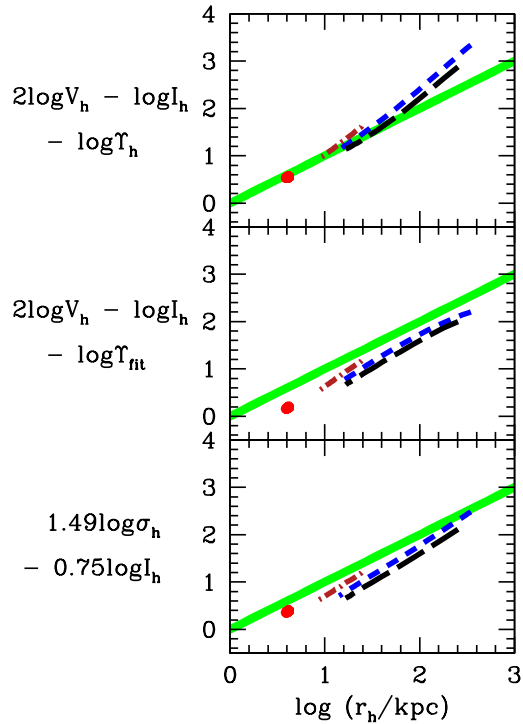}
\end{center}
\caption
{ Anisotropic rotator stellar models relative to the FM and FP. We plot their location relative to the TFM (using the known model value of $\Upsilon_h$) in the top panel,  the EFM (using the empirical fitting function for $\Upsilon_{h}$, $\Upsilon_{fit}$) in the middle panel, and the FP in the bottom panel. Model tracks represent results for $U_0 = -6$ (large red dot), $-8$ (dark red dot-dash), $-10$ (blue short dash), $-12$ (black long dash).  The 1:1 line represents the observed scaling relation in these axes and the width of the line matches the typically observed 0.1 dex scatter.  Along each model track, $\chi$ varies from 0.1 to 0.26, with smaller $\chi$ resulting in larger $r_h$.
For our shallowest potential, $U_0 = -6$, the models over this range of $\chi$ are effectively indistinguishable, so we designate the entire set with the large red dot.
Because the adopted $\Upsilon_*$ is arbitrary, we set $\Upsilon_*$ to a plausible value of 1. The models may be shifted vertically, by changing the adopted $\Upsilon_*$, to produce an optimal match.}
\label{fig:beta}
\end{figure}

We show where these models land on the FP and FM in Figure~\ref{fig:beta}. Each line type shows the results of varying a parameter, in this case $\chi$, for a particular choice of $U_0$.
Changes in $\chi$ produce sets of models that are much better aligned with either the FP or FM than the changes in $r_c$ we explored previously. For a fixed $r_c$, which these models have, changing $\chi$ mostly results in models sliding along the scaling relations.
Interestingly, the choice of $\chi$, and hence the division between systems dominated by rotational or pressure support, seems to have  relatively little effect on the structural aspects measured by these scaling relations.

We find a better correspondence between the scaling relations and models when we using the empirical scaling relations rather than the TFM.
This result may be surprising at first, but the empirical fits, either the FP or the EFM, provide a better match to the observations because hidden systematic structural behaviour (there are no $\Upsilon_h$ variations in these models because there is no dark matter and we adopt a constant $\Upsilon_*$) has been captured, and removed, in the empirical relations, demonstrating that $\Upsilon_{fit}$ is not entirely representative of $\Upsilon_h$ \citep[see also the ``weak homology" described by][]{bertin}.
If there is a surprise here, it is that the 
non-homologous behaviour produced in our simple models
accurately reproduces that 
of actual galaxies from previous studies \citep{bernardi,z12}.
This match bestows a degree of credibility on the models.

Before proceeding, we confirm that changes in $\zeta$ do not affect these conclusions.
We vary $\zeta$ from 0.01 to 0.21 and $\chi$ is fixed to be 0.133, which again is set to match that used in Wilson's Model A. We were unable (to the limits of our model gridding and integration steps) find acceptable models for the full range of $\zeta$ for all of our range of $U_0$. Acceptable models span ranges of  $\zeta$ that correspond to $0.36 \le v_h/\sigma_h \le 1.35$ for $U_0 = -12.0$, $0.285 \le v_h/\sigma_h \le 0.85$ for $U_0 = -10$,  $0.20 \le v_h/\sigma_h \le 0.46$ for $U_0 = -8$, and $0.16 \le v_h/\sigma_h \le 0.22$ for $U_0 = -6$.
The behaviour  of this set of models is qualitatively similar to that resulting from the variation of $\chi$,  demonstrating that the critical factor is $v_h/\sigma_h$.
Again we find that
the empirical fits, defined from real systems, for the FP and EFM, do a remarkable job of addressing the structural differences we find among our models, suggesting that those fits are not simply accounting for mass-to-light variations,  which do not exist in our star only models.

Finally, we return to two parameters that we set to be fixed in these models. First, we explore the dependence of our results on our choice of viewing angle, or inclination,
by examining a range of inclination from 15 to 75$^\circ$, where an inclination of 0 is defined to mean that  we measure the full rotation. As might be expected, changing the viewing angle results in only modest changes in $r_h$ (from 2.34 to 2.82 kpc in this model) and $\sigma$ (negligible), but a large change in $v_h$ from 87 to 27 km/s. However, because $V$ is the combination of $\sigma_h$ and $v_h$, the resulting effect of the changes in $v_h$ on where the models land relative to the scaling relations is mitigated. Following through on a specific example, we compare a case with low inclination and $v_h/\sigma_h = 1$ (for low $v_h/\sigma_h$ inclination does not matter), $V = \sqrt{87^2 + 87^2/2}$ to one with high inclination $V = \sqrt{87^2 + 27^2/2}$. The corresponding $2\log V$ terms differ by only 0.14, comparable to the observed scatter in the scaling relations. We conclude that our results are relatively insensitive to our choice of inclination. Second, we explore the dependence of the models on our choice of $r_c$. Changing our choice of fixed $r_c$ moves the model tracks, but does not change their orientation, which means that our freedom to move models vertically by altering $\Upsilon$ can remove the difference. 
We conclude that the model results are independent of our choice of $r_c$, to the degree that we are able to realistically manipulate $\Upsilon$.

\subsection{Models with stars and dark matter}
\label{sec:double}

We now add a second dynamical component to the models that is intended to represent the dark matter halo. In doing so we aim to have composite models that are broadly characteristic of galaxies and therefore match certain key aspects. We choose to match the ratio of dark matter to stars, $M_R$, and the ratio of dark matter half mass radius to the stellar half mass radius, $R_R$.  A variety of observations guide us, including observations of rotational curves \citep{begeman}, weak gravitational lensing of distant galaxies by foreground structure \citep{hoekstra,mandelbaum}, and dynamics of halo tracers \citep{z94,prada,more}. Considering one particular study as our guide \citep{lintott}, we converge on the loose criteria that $M_R > 10$.
For the constraint on $R_R$ we consider the range of half light radii of typical galaxy samples (1-10 kpc) and a range of NFW half mass radii for galaxies with $10^{10}$ --- 10$^{13}$ M$_\odot$ to
define a target range of $R_R > 5$

By construction, because we are adding the dark matter as an additional component described by the \cite{wilson} DF, our models do not match closely the dark matter mass density profiles found from cosmological simulations \citep[][hereafter NFW]{Navarro95}. Therefore, we focus on matching the mass density profile of the NFW model over the range of radii probed by the observable stellar dynamics. Deviations in the profile at the smallest radii are irrelevant for our purposes because there is little mass there. Deviations at the largest radii are also irrelevant because they are not probed by the stars within $r_h$.

\begin{figure}
\begin{center}
\includegraphics[width = 0.7 \textwidth]{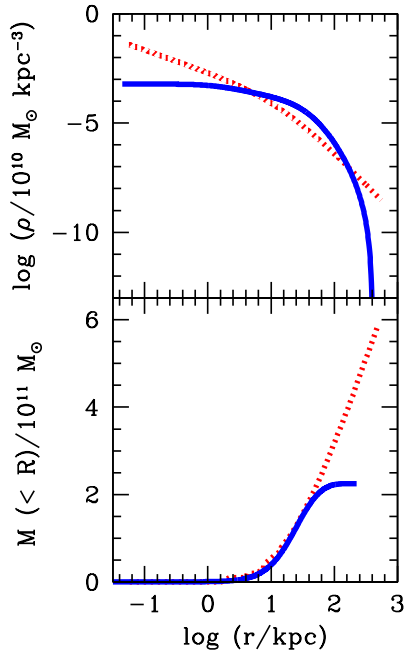}
\end{center}
\caption{Comparison of our baseline model dark matter density and enclosed mass profiles with those of an NFW halo. Our chosen model ($U_0=-9$, $\alpha_1 =$ 2.75, $\alpha_2$ = 0.4; solid line) is a poor match to a chosen NFW profile ($r_s =$ 7.7 kpc; dotted line). The profiles are normalised to each other at $r_h$ but are clearly quite different. In the lower panel we present the corresponding enclosed mass profiles, from which it becomes evident that differences only become substantial at radii beyond what is typically probed directly by observations. However, our models, if normalised to produce the enclosed mass at $r_h$, will underestimate the total mass out to the Virial radius by a factor of $\sim$ 2. }
\label{figure:density_profile}
\end{figure}

We provide a comparison of the density and enclosed mass profiles for what we adopt as our baseline model, obtained by setting $\chi = 0.133$, $\zeta = 0.067$ for the stellar component, $\chi = 0$, $\zeta = 0$ for the dark component, $U_0 = -9$ and $r_c = $ 2 kpc, to an NFW profile in Figure \ref{figure:density_profile}. Our model is evidently a poor match to the NFW profile, failing to reproduce the cusp-like central distribution
and the dark matter extent to large radii. However, the enclosed mass profiles are similar over about an order of magnitude in radius, out to a radius of about 60 kpc ($\log (r/{\rm kpc}) \approx 1.8)$.
For galaxies, $r_h$ is typically $<$ 10 kpc and although line-of-sight measurements will include stars at $r > r_h$ we expect a small fraction of the stars within a projected annulus of radius $r_h$ to have apocenteric distances as large as 60 kpc. In other words, the bulk of the stars that contribute to the kinematic measurement will lie within a radius where the enclosed dark matter mass is reasonably well approximated.
However, the total mass out to the Virial radius ($\sim$ few hundred kpc) in these models is likely to be low by a factor of $\sim$ 2 in comparison to a more accurate NFW dark matter profile. Because of this effect, we drop our target value of $M_R$, although not quite a factor of two to be conservative, to  $M_R > 7.5$.

\begin{figure}
\begin{center}
\includegraphics[width = 0.55 \textwidth]{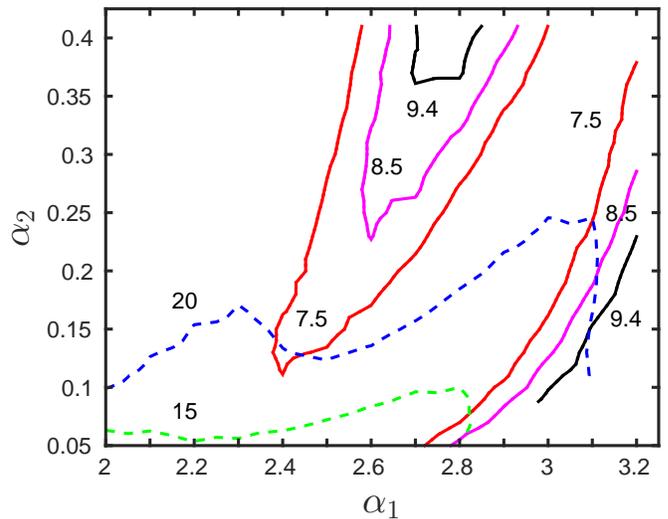}
\end{center}
\caption{Identifying models that match the properties of galaxies. We have produced baseline models with various combinations of $\alpha_1$, $\alpha_2$. We plot contours of the resulting $M_R$ and $R_R$. The bottom two dashed lines represent contours of $M_R$ (lower one for $M_R$ = 15, upper one for $M_R$ = 20). The other three contours (solid lines) represent $R_R$ = 7.5, 8.5 and 9.4, with the last one being represented by the smallest contour. The Figure demonstrates that in this set of models the choice of $\alpha_1$ mostly determines $R_R$ while that for $\alpha_2$ mostly determines $M_R$. }
\label{figure:contour}
\end{figure}

To probe the large parameter space of different star and dark matter distributions,
we calculate versions of our baseline model where we vary $n_i$, which affects the relative density of the $i$ components, and $\alpha_i$, which affects the relative energy among components.  From the resulting set of models, we highlight those that satisfy our $M_R$ and $R_R$ criteria and therefore appear most similar to actual galaxies. We present the dependence of $M_R$ and $R_R$ on the choice of $\alpha_1$ and $\alpha_2$ in Figure \ref{figure:contour}.
The contours in the Figure \ref{figure:contour} represent constant values of $M_R$ and $R_R$ obtained with various combinations of  $\alpha_1$ and $\alpha_2$ (in this set of models we set $n_1$ to 50000, $n_2$ is set to be 0.01). For example, $R_R$ is $>$ 9 when $\alpha_1 \approx 2.75$, $\alpha_2 > 0.325$. The full range of models explored, regardless of whether they produce $M_R$ and $R_R$ that we consider to be rough matches to real galaxies, produces galaxies that span $-0.5 \le \log(r_h/{\rm kpc}) < 2$ and $0.1 < v_h/\sigma_h < 0.8$. Over this full set of models, the smallest and largest values of $\sigma_h$ differ by 27\%.

\begin{figure}
\begin{center}
\includegraphics[width = 0.7 \textwidth]{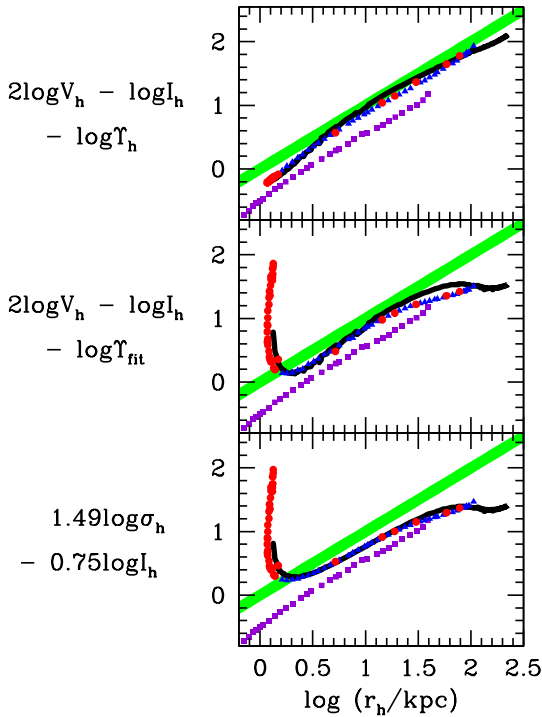}
\end{center}
\caption{Dark + stellar models relative to the FM and FP.
Models span a range of choices for $\alpha_1$ and $\alpha_2$, resulting in different values of $M_R$ and $R_R$.
We plot their location relative to the TFM (using the known model value of $\Upsilon_h$) in the top panel,  the EFM (using the empirical fitting function for $\Upsilon_{h}$, $\Upsilon_{fit}$) in the middle panel, and the FP in the bottom panel.
The 1:1 line represents the observed scaling relation in these axes and the width of the line matches the typically observed 0.1 dex scatter.
The blue triangles represent those models defined as acceptable because they broadly match the properties of actual galaxies. Red circles represent those models that do not match. In both cases, adopting $r_c$ = 2 kpc. The purple squares represents all models adopting $r_c$ = 1 kpc. We remind the reader that vertical offsets from observed relation can be removed with a judicious choice of $\Upsilon$ and so are not to be considered as discriminating. The black solid line represents models of using another dimensionless parameter set ($U_0 = -12.0, \chi = 0.06$ and $\zeta = 0.02$) and $r_c = 2$ kpc.
The success of the TFM illustrates that the inclusion of dark matter does not violate the structural homology presumed in the FM formalism. The failure of the EFM and FP, particularly at small $r_h$ for $r_c = 2$ kpc, demonstrates that the empirical fits used in these relations cannot be extrapolated beyond the parameter range used to calibrate those relations.}
\label{figure:DM_FM}
\end{figure}

All of our baseline models, over the range of $\alpha_1$ and $\alpha_2$ sampled, are plotted in Figure~\ref{figure:DM_FM} and span $1.7 < R_R < 10$ and $2.8 < M_R < 1000$. Because of the introduction of dark matter (DM), $\Upsilon_h$ is no longer equivalent to $\Upsilon_*$ but is instead
\begin{equation}
\Upsilon_h = \frac{M_h^* + M_h^{DM}}{M_h^*}\times \Upsilon_*
\end{equation}
where $M_h^*$ is the projected stellar mass within $r_h$, $M_h^{DM}$ is the projected DM mass within the same radius, and $\Upsilon_*$ is, as before, our adopted stellar mass-to-light ratio, which we have set to 1 and consider to be a constant. We do not change the empirical fitting formula for $\Upsilon_h$, $\Upsilon_{fit}$, because it was calculated using real systems that contain DM.

In Figure~\ref{figure:DM_FM} we show that the full range of our baseline models for our standard adoption of $r_c=2$ kpc and for $r_c=1$ kpc. Even though some of these models are well beyond the range of parameters of real galaxies, they fit very well on the TFM. Because structural homology is built into the TFM and because the models lie tightly along the relation, we conclude that the inclusion of dark matter does not introduce significant deviations from structural homology. In contrast, the gross systematic failures of the EFM and FP for both small and large $r_h$ systems, for $r_c=2$ kpc, demonstrate the importance of the inclusion of an accurate estimate of the mass-to-light ratio term in those relations. The value of $\Upsilon_h$ is implicit in the FP and explicit in the EFM, but both fail because they depend on empirical fits to a set of actual galaxies that do not extend over the full parameter range of our model systems. For example, our smallest $r_h$ models still have high $\sigma_h$ ($\sim 150$ km/sec) and so do not match actual galaxies. At large $r_h$ we may be seeing a slightly different phenomenon in that there is some empirical evidence that the largest galaxies do deviate from the empirical scaling relations \citep{bernardi}. The relative success of the TFM in contrast to the failure of the FP and EFM in this context does not contradict the reverse result we found previously when exploring the stellar models by varying $U_0$ and $r_c$. In those cases, changes in $U_0$ and $r_c$ led to structural changes that we proposed were being implicitly modelled out by the empirical scaling relations. Here, we have so far only probed the effects of changes in the relative contribution and distributions of luminous and dark matter, for a model with fixed $U_0$ and $r_c$.

As an aside, because scaling relations can be rewritten as mass estimators, we note that both the EFM and FP would produce artificially large values of $\Upsilon_h$ for some models with the smallest $r_h$. This deviation is reminiscent of the situation for the faintest dwarf galaxies, in that standard mass estimators result in large values of $\Upsilon$, but in those cases values have been vetted using dynamical modeling \citep{walker}. Here, our smallest galaxies are not galaxies found in nature, but the cautionary note still stands in extrapolating simple mass estimators beyond parameters ranges where those estimators have been validated with detailed studies.

\begin{figure}
\begin{center}
\includegraphics[width = 0.7 \textwidth]{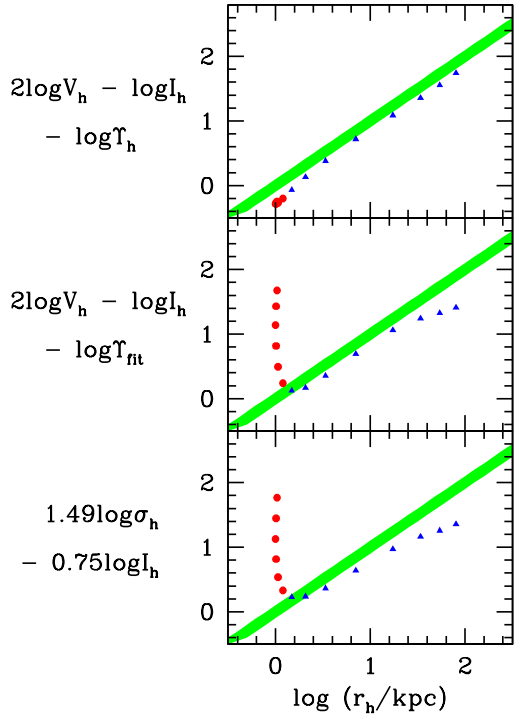}
\end{center}
\caption{Dark + stellar models relative to the FM and FP.
Models span  $-5 < U_0 < -12$ in steps of $\Delta U_0 = 0.5$, with all other parameters remaining fixed.
We plot their location relative to the TFM (using the known model value of $\Upsilon_h$) in the top panel,  the EFM (using the empirical fitting function for $\Upsilon_{h}$, $\Upsilon_{fit}$) in the middle panel, and the FP in the bottom panel.
The blue triangles represent what we have defined as acceptable models in that they match broadly the properties of actual galaxies and the red circles represent models that do not match those properties.
The 1:1 line represents the observed scaling relation in these axes and the width of the line matches the typically observed 0.1 dex scatter.
The models may be shifted vertically, by changing the adopted $\Upsilon_*$, to produce an optimal match.
With the exception of some models with small $r_h$ that lie beyond the parameter range used to obtain the empirical relations, changes in $U_0$ cause the models to slide nearly parallel to the relations.
}
\label{figure:DM_u0}
\end{figure}

\begin{figure}
\begin{center}
\includegraphics[width = 0.7 \textwidth]{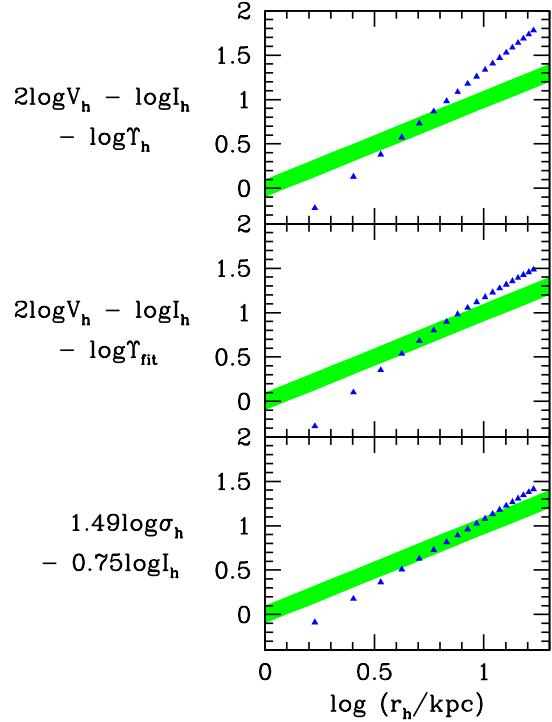}
\end{center}
\caption{
Dark + stellar models relative to the FM and FP.
Models span  $1 < r_c < 10$ kpc in steps of $\Delta r_c = 0.5$ kpc, with all other parameters remaining fixed.
We plot their location relative to the TFM (using the known model value of $\Upsilon_h$) in the top panel,  the EFM (using the empirical fitting function for $\Upsilon_{h}$, $\Upsilon_{fit}$) in the middle panel, and the FP in the bottom panel.
The blue triangles represent what we have defined as acceptable models in that they match broadly the properties of actual galaxies and the red circles represent models that do not match those properties. There are no red circles, demonstrating the choice of $r_c$ does not influence $M_R$ and $R_R$.
The 1:1 line represents the observed scaling relation in these axes and the width of the line matches the typically observed 0.1 dex scatter.
The models may be shifted vertically, by changing the adopted $\Upsilon_*$, to produce an optimal match.
We find that varying $r_c$ moves model across the relation, demonstrating that scatter in $r_c$ would be reflected as increased scatter in the scaling relations.
}
\label{figure:DM_rc}
\end{figure}

When we do vary $U_0$ or $r_c$ and fix our other parameters (Figures \ref{figure:DM_u0} and \ref{figure:DM_rc}) we confirm the tendencies noted in the original non-rotating stellar models. Varying $U_0$ causes the models to move nearly in parallel to the scaling relations. The only exceptions to this statement are found among the extreme models that do not match observed galaxies and for which the empirical relations are not expected to work. In contrast, varying $r_c$ moves the systems across the scaling relations. These results confirm that the key to understanding the scaling relations lies in understanding the physics that constrain our choices of $r_c$.

As one further test of these results, we also briefly explore the consequences of varying multiple dimensionless parameters simultaneously. We have in general avoided varying more than one parameter because of the added difficulty in interpreting the results, but one might suspect that the models could behave differently when multiple parameters are varied in concert. To check this, we adopt another of Wilson's models ($U_0 = -12.0, \chi = 0.06$ and $\zeta = 0.02$). We choose this particular model because from those presented in his Table 3 it is among those with the largest rotational support and anisotropy, in starkest contrast with the baseline model we have mostly considered. We then again vary the normalisation and radial scaling of the dark matter component relative to the stellar component and the resulting models are included in Figure \ref{figure:DM_FM}. We have kept $r_c$ the same (2 kpc), and the models are nearly indistinguishable from our previous models. This result demonstrates that degree to which varying internal structural and kinematic parameters fails to cause significant variations in the scaling relations, even when multiple parameters are varied and the internal structure is markedly different.

The finding of a consistent central property among galaxies, $r_c$, is reminiscent of measurements of a constant enclosed mass or central surface density within a fixed physical radius \citep{strigari,donato,salucci, burkert}. Specifically, \cite{donato}, \cite{salucci} and \cite{burkert} find that $r_c \rho_0$ is a constant. In our models, $\rho_0$ scales with our choice for the velocity scale normalisation, $v_*$, and radial scale normalisation, $r_*$, through the scaling term $\rho_{0*} = (9/4\pi G)v_*^2/r_*^2$. The Wilson models are initialised such that $r_* = r_c$. Therefore, the scaling of the combination of $r_c \rho_0$ can be expressed as $(9/4\pi G) v_*^2/r_c$. In the models described so far, we did not vary $v_*$. Therefore, if we now impose a requirement that $r_c \rho_0$ be a constant, it is equivalent to requiring that $r_c$ be constant --- which is what we found necessary to reproduce the scaling relations. We now examine how allowing changes in $v_*$ affects this constraint on $r_c$.

Changes in $r_c$ and $v_*$ only cause scalings of the model quantities, not a change in the density or velocity profiles when those are expressed in model units. To consider how a model will migrate in the FM plot we need to determine how rescaling affects the model's position in both $x$ and $y$.
The scaling along the $x$-axis is simple in that $r_h$ will scale directly with the choice of $r_*$. The scaling along the $y$-axis is slightly less straightforward.
In the case of the TFM, the $y$-axis is $2\log V - \log I_h - \log \Upsilon_h - C_{FM}$. The last term is manifestly a constant and so does not affect anything other than the normalisation, which we leave free to vary. If we consider stellar only systems, then $\Upsilon_h \equiv \Upsilon_*$ and so this term too is a constant. The first term of the expression will evidently scale with $v_*^2$ because it can be written as $\log V^2$.  The second term has a dependence of mass/radius$^2$, which itself scales as $v_*^2 r_*/r_*^2$ or finally $v_*^2/r_*$. Because we showed previously that the empirical constraint of $r_c\rho_0 = $ constant is equivalent to requiring $v_*^2/r_c$ to be a constant, this term is itself a constant. Therefore, the requirement that $v_*^2/r_c$ is a constant, which comes from requiring that $r_c\rho_0$ be constant, results in $2\log V - \log r_h = $ constant, which corresponds to a slope = 1 line in the FM space. Therefore, changes in $v_*$ and $r_c$ that maintain $r_c\rho_0 = $ constant will slide points along the FM. Alternatively, we could have begun this discussion by requiring the existence of the FM and recovering the constraint that $r_c \rho_0 = $ constant. In either way, we see that the two results are related. The argument does not hold as neatly for the FP, but the FP does not describe the properties of the dwarf spheroidal galaxies for which the $r_c \rho_0 =$ constant property was first noticed and so is not as general.

Although this discussion has demonstrated that requiring our models match the global galaxy scaling relations is equivalent to the finding $r_c \rho_0$ = constant, it has not provided a physical explanation for either the existence of the scaling relations or the constancy of $r_c \rho_0$.  It is worth noting, however, that these two apparently independent constraints, are in fact dependent and may point to the same physical cause.

\section{Conclusions}
\label{sec:summary}

We have utilised dynamical models of early type galaxies first presented by \cite{wilson} to explore the origin of some common scaling relations. In particular, these are simple dynamical models that include rotation  and anisotropic velocity dispersions, important characteristics for many elliptical galaxies, and that incorporate multiple dynamical components, a requirement so that we can include dark matter in addition to stars. The distribution function defined by these models is self-consistent in that it gives rise to the host gravitational potential. Given the distribution function, we calculate the necessary observables including the half light radius, $r_h$, the velocity dispersion and rotation velocity at the half light radius, $\sigma_h$ and $v_h$, and the mean surface brightness within the half light radius, $I_h$. These parameters then allow us to place these models in the context of the Fundamental Plane \cite[FP;][]{djorgovski,dressler} and Manifold \cite[FM;][]{z06}. We begin our discussion with our simplest models, single component with no rotation  and isotropic velocity dispersions, and continue to our most complicated, containing dark matter and stars and where the stars are differentially rotating  and have anisotropic velocity dispersions.

We investigate the effect of varying a set of parameters including the  physical scale of the scaling radius ($r_c$), the central potential ($U_0$), the rotation speed ($\chi$), parameter that quantify the degree of differential rotation and anisotropy ($\chi$ and $\zeta$), and the relative masses and concentrations of stars vs. dark matter. As we vary each parameter in turn, we find that changing $r_c$  produces the results that are most discrepant with the existence of the scaling relations. Once $r_c$ is fixed, changing any of the other available parameters principally moves models along the scaling relations. Because $r_c$ acts as a scaling parameter, the particular choice of $r_c$ does not affect the density or velocity profiles in model units, we find that intrinsically the models reproduce the scaling relations well.

Among the models that include dark matter, we have some whose characteristics, such as the mass ratio between dark matter and stars, do not match actual galaxies. Because these systems are beyond the range of properties used to produce empirical scaling relations such as the FP and FM, these systems can fail to satisfy the empirical scaling relations. However, the inclusion of the correct mass-to-light ratio at the half light radius, $\Upsilon_h$, which we can calculate directly from the models, places the system once again back on the FM. This result  confirms the near-homology of the set of model galaxies and the importance of having the appropriate approximation for $\Upsilon_h$. However, it is quite remarkable that over the range of systems that do have real-world analogs the empirical relations are reproduced well by these simple models  with no adjustment other than requiring a fixed value of $r_c$.

Finally, presuming that nature does allow for a large range of $r_c$, we explore the connection between our results and the empirical finding that $r_c \rho_0$ is a constant for galaxies \citep{donato}.  We find that this empirical finding is equivalent, for our models, to demanding that the models produce a scaling relation consistent with the FM. As such, the two unrelated observational results are connected. In other words, a requirement that the models reproduce a scaling relation would produce $r_c \rho_0 = $ constant and conversely a requirement that $r_c \rho_0 = $ constant would produce the scaling relations. Which is the driver, and what the physical mechanism is for imposing it, remain open questions. While there are theoretical efforts to address some of these questions \citep[see for example][]{governato}, we show here that these are not independent constraints and that achieving one may well produce the other.

\section{acknowledgments}
DZ acknowledges financial support from NASA ADAP NNX12AE27G and NSF grant AST-1311326. We thank our diligent referee for providing comments that led to significant improvements of this manuscript.
\bibliography{bibliography}

\appendix

\section{Testing the calculations}

We now discuss three ways we confirm the reliability and robustness of our models. We present the convergence of our numerical schemes with regards to how we define our spatial grid and integration step size, calculate the concurrence of our models with Virial equilibrium, and compare our results to those presented by \cite{wilson}.

Implementation of any numerical scheme involves a variety of choices regarding the resolution of the radial and azimuthal grids, and the integration step size in our numerous numerical integrations. We show in Table~\ref{table:Variables} the results of one set of our exploration of the dependence of our results on these choices. For a model with the parameters of Wilson's (1975) Model A, we show how values of the tidal radius in model scale units, $r_t$, the potential at infinite radius in model units, $U_\infty$, and the stellar mass, $M_*$, depend on the choice of the number of radial bins, $r_N$, azimuthal bins, $\theta_N$, and integration steps, I$_N$. Based on such tests, we set our standard models to have $r_N = 750$, $\theta_N = 51$, and I$_N = 1000$.  The consistency among results for values of these parameters larger than the chosen ones demonstrates that we have achieved convergence for of our grid definition and integration conventions, at least for this particular model. Corresponding tests spanning our base parameter set confirm our choices.

For the second of our tests, we evaluate
the ratio of the kinetic energy, $K$, to the potential energy, $W$. This ratio approaches the theoretically expected value of 0.5 as the precision of our modelling increases.
We find that higher resolution does not alter the results nor does it improve the approximation to Virial equilibrium. Based on extensive testing, our technical choices appear to be appropriate for a wide range of models, but are not necessarily appropriate for all models. For example, models that deviate significantly  more from spherical symmetry than does this model will require a larger value of $\theta_N$ to properly sample this asymmetry. In practice, if Virial equilibrium is not confirmed to the level of precision seen for the best models in Table 1, we revisit our choices and recalculate the model or reject the model. Although in Table \ref{table:Variables} we present the Virial criterion for a star-only model, this criteria is also satisfied for star + dark matter models. We only accept models if the Virial ratio is $0.5000 \pm$0.0014.

\begin{figure}
\begin{center}
\includegraphics[width = 0.48 \textwidth]{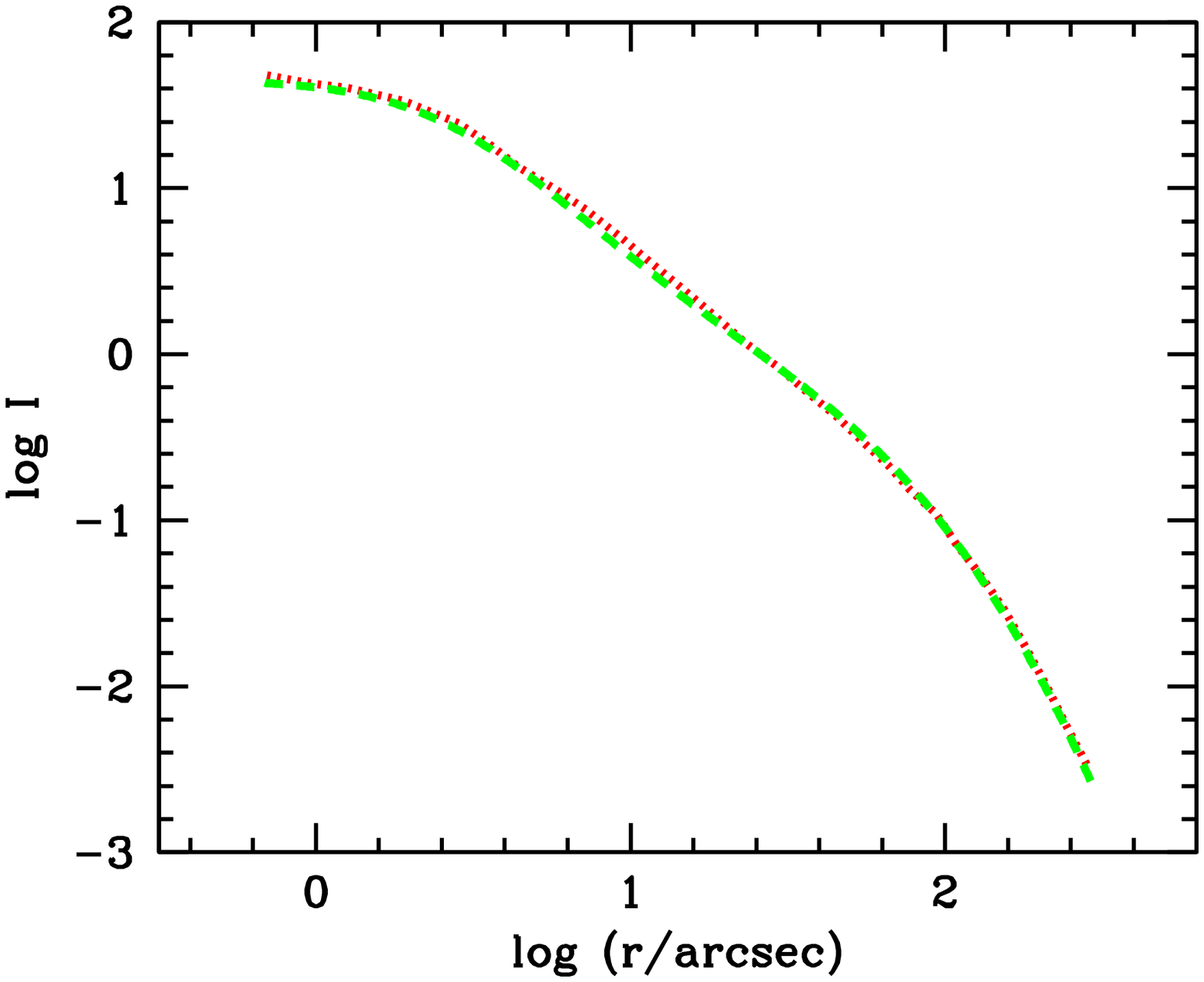}
\includegraphics[width = 0.48 \textwidth]{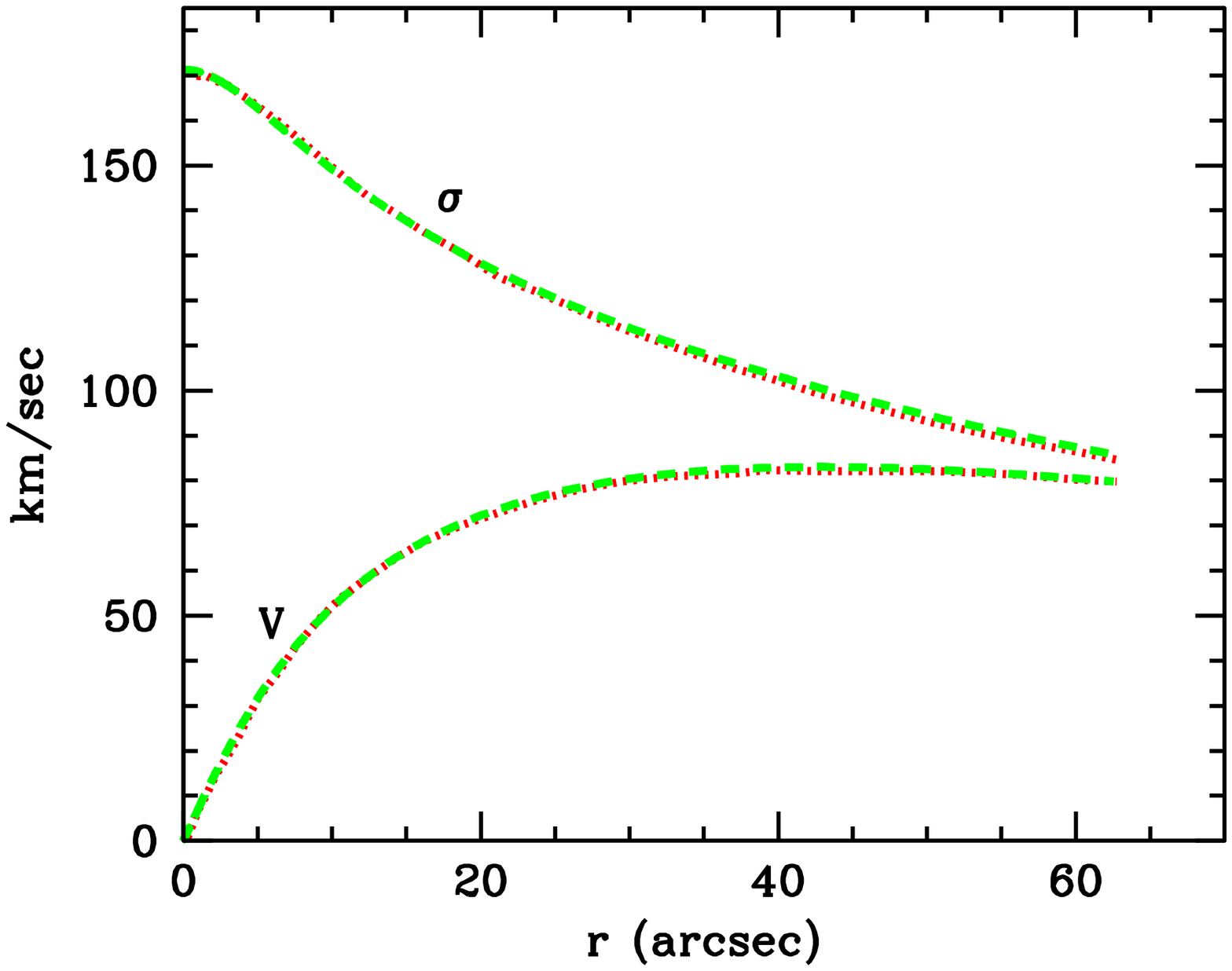}
\end{center}
\caption{
Reproduction of Wilson's (1975) Model A results using our calculations. In the upper panel we show the surface brightness profile ($\log I$ vs. $\log r$), where $I$ is in units of one
20th mag star per square arcsec and $r$ is in units of arcsec. In the
lower panel we show the projected radial velocity and velocity dispersion versus radius. The Wilson (1975) data are digitised from a plot in his paper and shown as the red dotted line. Our calculations, shown as the green dashed line, match well.}
\label{Figure:rho_compare}
\end{figure}

Finally, we compare our results with those of \cite{wilson} for the first of his three models, Model A, which he constructed as representative of NGC 3379. In Figure \ref{Figure:rho_compare} we show the result of the comparison for the projected stellar density, as converted into surface brightness, and for the line-of-sight velocity dispersion and rotation velocity profiles. For this comparison, we digitised the corresponding plots from \cite{wilson} and include those results in our comparison. The agreement is manifestly excellent.

\begin{table}
\caption{Model Convergence}
\begin{tabular}{rrrcccr}
\hline
$r_N$&$\theta_N$&I$_N$&$\log(r_t)$&$U_\infty$&$M_*$&$K/|W|$\\
&&&&&[$10^{10} M_\odot$]\\
\hline
 50 & 25 & 50 & 2.61 & 0.124 & 9.077 & 0.5100 \\
 50 & 25 & 250 & 2.61 & 0.124 & 9.077 & 0.5089 \\
 100 & 25 & 250 & 2.65 & 0.143 & 7.860 & 0.5019 \\
 250 & 51 & 250 & 2.61 & 0.162 & 7.670 & 0.5014 \\
 500 & 51 & 500 & 2.57 & 0.170 & 7.360 & 0.4993 \\
 750 & 51 & 1000 & 2.57 & 0.170 & 7.355 & 0.4997 \\
 1000 & 75 & 1000 & 2.57 & 0.170 & 7.350 & 0.4997 \\
 \hline
\label{table:Variables}
\end{tabular}
\end{table}

\end{document}